\documentclass[conference]{IEEEtran}
\IEEEoverridecommandlockouts

\usepackage{cite}
\usepackage{amsmath,amssymb,amsfonts}
\usepackage{algorithmic}
\usepackage{graphicx}
\usepackage{textcomp}
\usepackage{xcolor}
\usepackage{cite}
\usepackage{amsmath,amssymb,amsfonts}
\usepackage{algorithmic}
\usepackage{graphicx}
\usepackage{textcomp}
\usepackage{xcolor}
\usepackage{tikz}
\usepackage{amsmath}
\usepackage{booktabs}
\usepackage{subcaption}
\usepackage{siunitx}
\usepackage{threeparttable}

\def\BibTeX{{\rm B\kern-.05em{\sc i\kern-.025em b}\kern-.08em
    T\kern-.1667em\lower.7ex\hbox{E}\kern-.125emX}}
\begin{document}

\title{PropSplat: Map-Free RF Field Reconstruction via 3D Gaussian Propagation Splatting

\thanks{This research was supported in part by National Science Foundation (NSF) grant OAC-2512931.

© 2026 IEEE. Personal use of this material is permitted. Permission from
IEEE must be obtained for all other uses, in any current or future media,
including reprinting/republishing this material for advertising or promotional
purposes, creating new collective works, for resale or redistribution to servers
or lists, or reuse of any copyrighted component of this work in other works.}
}

\author{
    \IEEEauthorblockN{
    William Bjorndahl\IEEEauthorrefmark{1},
    Maninder Pal Singh\IEEEauthorrefmark{2}, Farhad Nouri\IEEEauthorrefmark{1}, and Joseph Camp\IEEEauthorrefmark{1}}
    \IEEEauthorblockA{\IEEEauthorrefmark{1}Department of Electrical and Computer Engineering, Southern Methodist University, Dallas, TX, USA}
    \IEEEauthorblockA{\IEEEauthorrefmark{2}Department of Engineering Technology, University of Houston, Houston, TX, USA}
    Email: \{wbjorndahl, fnouri, camp\}@smu.edu, msingh35@central.uh.edu
}

\maketitle

\begin{abstract}
    Building a site-specific propagation model typically requires either ray-tracing over detailed 3D maps or dense measurement campaigns. Both approaches are expensive and often infeasible for rapid deployments where geographic data is unavailable or outdated. We present PropSplat, a map-free propagation modeling method that reconstructs radio frequency (RF) fields using 3D anisotropic Gaussian primitives. Each Gaussian encodes a scalar path loss offset relative to an explicit baseline path loss model with a learnable path loss exponent. Gaussians are initialized along observed transmitter--receiver paths and optimized end-to-end to learn the propagation environment without external information like floor plans, terrain databases, or clutter data. We evaluate PropSplat against wireless radiance field methods NeRF$^2$, GSRF, and WRF-GS+ on two real-world datasets. On large-scale outdoor drive-tests spanning multiple topographical regions at six sub-6\,GHz frequencies, PropSplat achieves 5.38\,dB RMSE when training measurements are spaced 300\,m apart and outperforms WRF-GS+ (5.87\,dB), GSRF (7.46\,dB), and NeRF$^2$ (14.76\,dB). On indoor Bluetooth Low Energy measurements, PropSplat achieves 0.19\,m mean localization error, an order of magnitude better than NeRF$^2$ (1.84\,m), while achieving near-identical received signal strength prediction accuracy. These results show that accurate site-specific propagation reconstruction is achievable from sparse RF-native measurements. The need for geographic data as a prerequisite for scalable RF environment modeling is reduced.
\end{abstract}

\begin{IEEEkeywords}
RF field reconstruction, 3D Gaussian splatting, sparse measurements, path loss prediction, machine learning

\end{IEEEkeywords}

\section{Introduction}

Path loss prediction estimates signal attenuation between transmitter (Tx) and receiver (Rx) locations. Tasks like coverage planning, interference management, and deployment optimization rely on accurate path loss predictions. As networks densify and operate across wider bandwidths and higher frequencies, operators increasingly need site-specific RF field reconstruction that can be refreshed quickly and evaluated at scale. Yet today’s high-accuracy options typically demand either (i) computationally intensive physics-based simulation with detailed 3D environment models, or (ii) extensive measurement campaigns, often combined with rich geographic priors.

This dependence on geographic information system (GIS) assets is a practical barrier. Terrain models, building databases, clutter maps, imagery, and LiDAR can be expensive to acquire and process, and may be outdated, incomplete, or unavailable. These constraints limit propagation knowledge in rapid deployment scenarios, e.g., temporary events or emergencies, newly developed regions, and dynamic environments where construction, vegetation, and other changes can quickly invalidate map priors. At the same time, relying purely on dense drive testing is costly and often infeasible to conduct frequently. These realities motivate the question: \emph{can we reconstruct accurate RF fields without any geographic priors, using sparse wireless measurements?}

We answer this question with PropSplat, a map-free propagation modeling approach inspired by 3D Gaussian Splatting (3DGS) from computer graphics \cite{10.1145/3592433}. PropSplat represents an environment as a set of learnable 3D anisotropic Gaussian primitives, each carrying a scalar offset relative to a log-distance baseline path loss model with a learnable path loss exponent. Gaussians are initialized along Tx--Rx measurement paths and optimized end-to-end so that, when projected onto a Tx--Rx link, their spatially and directionally weighted contributions sum to explain the offset between measured and baseline-predicted attenuation. The Gaussians do not explicitly represent geometry. They are empirical, spatially-localized corrections whose aggregate effect captures whatever site-specific propagation phenomena are present in the measurements.

PropSplat makes two primary contributions:
\begin{itemize}
    \item \textbf{Map-free RF field reconstruction:} PropSplat uses Tx location, Rx location, frequency, and path loss/received signal strength indicator (RSSI) measurements. This map-free method enables deployment when geographic priors are unavailable or unreliable.
    \item \textbf{High accuracy with ultra-sparse measurements:} By learning path loss offsets with an explicit 3D Gaussian representation, PropSplat trains effectively with data spaced far apart, e.g., widely spaced outdoor drive-test samples, while supporting fast inference that is suitable for large-scale evaluation.
\end{itemize}

\section{Related Work}
\label{sec:related_work}

RF field reconstruction spans empirical/physics-based modeling, statistical interpolation, and learning-based methods. Since this paper targets map-free reconstruction from sparse measurements, we focus on three relevant lines of work: (i) wireless radiance fields adapted from computer graphics rendering, (ii) machine learning using map and geographic features, and (iii) map-free statistical methods.

\subsection{Wireless Radiance Fields}

Neural rendering methods such as NeRF represent scenes with continuous volumetric functions learned by multilayer perceptrons (MLPs)~\cite{10.1145/3503250}, while 3DGS replaces implicit fields with explicit anisotropic Gaussians for efficient optimization and rendering~\cite{10.1145/3592433}. Several RF works adapt these ideas. NeRF$^2$ models volumetric RF properties and can capture complex interactions, but typically requires long training times~\cite{10.1145/3570361.3592527}. GSRF improves efficiency by adopting 3DGS-style primitives with complex-valued Fourier--Legendre expansion coefficients that are directly optimized per
  Gaussian, eliminating MLP regressors entirely while modeling directional RF effects~\cite{yang2025gsrf}.
  Other approaches pursue uncertainty-aware or active sampling with Gaussian-process-based formulations~\cite{gau2024active}, and multi-modal fusion methods that leverage visual priors when available (e.g., camera-derived geometry)~\cite{10.1145/3666025.3699351}. WRF-GS and WRF-GS+ adapt 3D Gaussian splatting to wireless radiation field reconstruction by treating Gaussian primitives as virtual Txs that carry learnable complex signal and attenuation attributes, enabling explicit visualization and synthesis of channel characteristics~\cite{11044513}, \cite{11258087}.

Collectively, wireless radiance field methods demonstrate scene-style representations for channel characterization. However, they often require heavier parameterizations than needed for sparse, map-free path loss reconstruction.

\subsection{Machine Learning with Geographic Features}

A large body of work predicts path loss by extracting features from GIS assets such as digital terrain/surface models, building/clutter databases, imagery, and LiDAR. Examples include inferring cellular key performance indicators by combining geographical features from LiDAR with crowdsourced measurements using neural networks~\cite{9384266}, leveraging path-profile obstacle depth from terrain/surface models \cite{ethier2024machine}, and using deep or tree-based predictors from terrain and clutter-derived inputs \cite{dempsey2025map, bocus2025application}. While these approaches can be accurate, their performance and portability depend on the availability, resolution, and freshness of the underlying geographic priors---and on nontrivial preprocessing pipelines.

\subsection{Map-Free Statistical and Data-Driven Methods}

Without map priors, classical interpolation and probabilistic methods like Kriging and Gaussian processes can estimate path loss fields from measurements, sometimes augmented with shadowing models or dimensionality reduction \cite{s20071927, 8598897}. These methods can provide uncertainty estimates but often struggle with sharp spatial transitions and scale poorly with large datasets due to unfavorable computational complexity. Moreover, many data-driven map-free approaches still require dense measurements to reach low error in environments that contain much clutter.

\subsection{PropSplat Position}

PropSplat is a wireless radiance field approach and is especially beneficial for RF field reconstruction under operations with ultra-sparse measurements. Rather than modeling the full RF field with an implicit network, PropSplat learns a baseline distance-dependent trend plus a set of explicit 3D Gaussians that explain site-specific offsets. Initializing Gaussians along observed Tx--Rx paths constrains learning to regions supported by measurements, improving data efficiency. Anisotropic primitives capture directional propagation effects without requiring external geometry. This combination enables scalable inference and accurate reconstruction in scenarios where obtaining dense measurements or geographic priors is impractical.

\section{PropSplat Methodology: 3D Gaussian Propagation Splatting}
\label{sec:gaussian_representation}

\begin{figure}[t]
    \centering
    \includegraphics[width=1\linewidth]{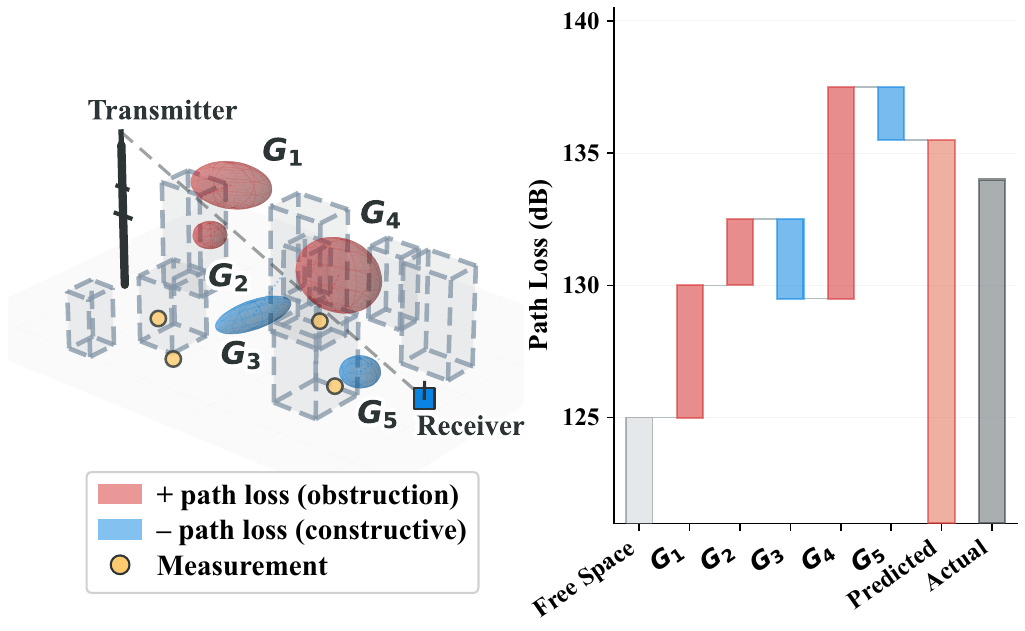}
    \caption{PropSplat represents \textit{radio propagation effects} as learnable 3D Gaussian ellipsoids \textit{splatted} throughout the environment. Each Gaussian encodes a spatially localized offset from a baseline path loss model. Red Gaussians ($G_1$, $G_2$, $G_4$) represent increased path loss due to obstruction or shadowing and blue Gaussians ($G_3$, $G_5$) represent decreased path loss from constructive propagation effects such as reflection. Buildings (dashed outlines) are shown for context only as PropSplat requires no map, building, or terrain data. (Right) A single path loss prediction is formed by summing each Gaussian's weighted contribution over the free-space baseline, closely matching the actual measured path loss.}
   
    \label{fig:concept}
\end{figure}

\begin{figure*}[t]
    \centering
    \includegraphics[width=1\linewidth]{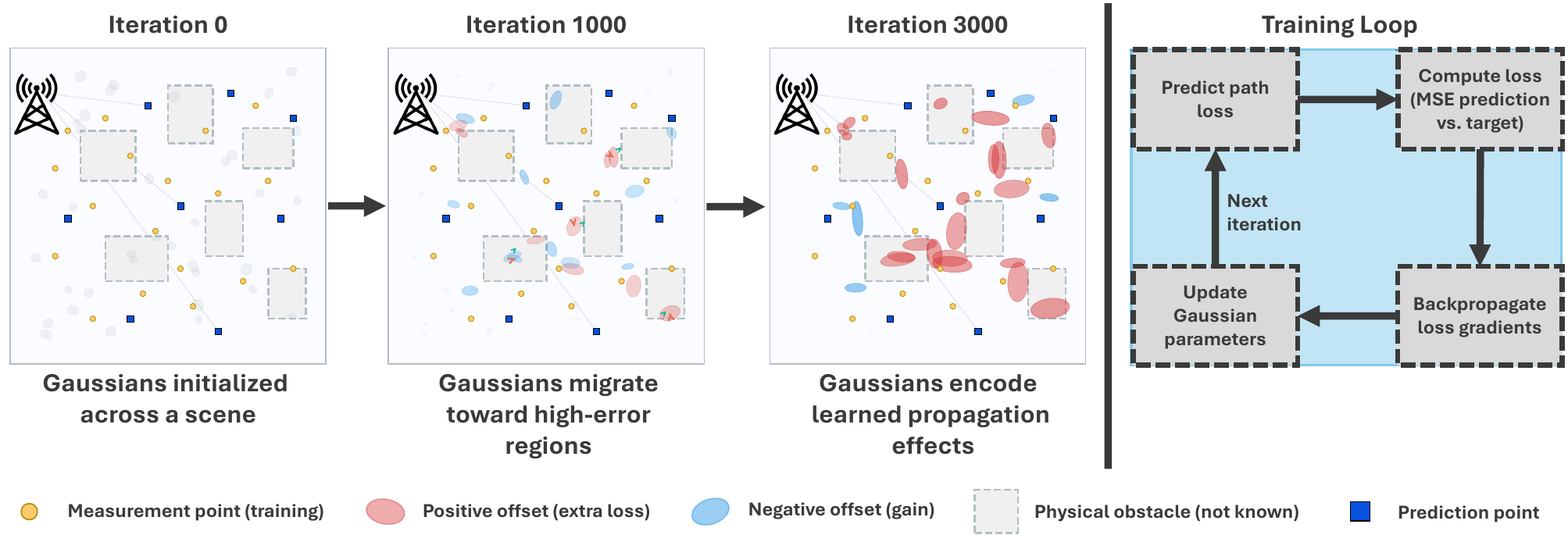}
    \caption{PropSplat training process where Gaussians evolve from initialization in Iteration~0 through migration toward high-error regions and eventual convergence around learned propagation effects. The training loop predicts path loss using the current Gaussian parameters and learned path loss exponent, computes mean squared error (MSE) loss against measured data, backpropagates gradients to all learnable parameters (position, scale, rotation, offset), and updates them via Adam optimization.}

    \label{fig:training}
\end{figure*}

Our proposed method models the RF propagation environment using a collection of 3D Gaussians. These Gaussians do not directly represent physical objects. They instead learn to represent the impact of the physical environment on the RF signal path loss relative to a baseline free-space model. This concept is illustrated in Fig.~\ref{fig:concept}. The parameters of these Gaussians are optimized to best explain the observed path loss measurements from a given dataset. Each Gaussian $G_i$ in our representation is defined by a set of learnable parameters $\Theta_i$:
\begin{equation}
    \Theta_i = \{\mu_i, s_i, q_i, o_i\}
\end{equation}

where 
\begin{itemize}
    \item $\boldsymbol{\mu}_i \in \mathbb{R}^3$: The mean or center position of the Gaussian in 3D space.
    \item $\mathbf{s}_i \in \mathbb{R}^3$: A vector of scaling factors along the Gaussian's local axes. This allows for anisotropic shapes, modeling direction-dependent effects. For numerical stability during optimization, we parameterize this using the logarithm, $\log(\mathbf{s}_i)$.
    \item $\mathbf{q}_i \in \mathbb{R}^4$: A normalized quaternion ($||\mathbf{q}_i||_2 = 1$) representing the rotation or orientation of the Gaussian's local axes relative to the world coordinate system.
    \item $o_i \in \mathbb{R}$: A scalar value representing the peak path loss offset contribution (in dB) associated with this Gaussian. A positive $o_i$ indicates that this Gaussian tends to increase the path loss, e.g., modeling shadowing, while a negative $o_i$ suggests a decrease, potentially modeling constructive interference or reflections.
\end{itemize}

The scale vector $\mathbf{s}_i$ and rotation quaternion $\mathbf{q}_i$ together define the 3D shape and orientation of the Gaussian. They can be used to construct the covariance matrix $\boldsymbol{\Sigma}_i$ for the $i$-th Gaussian:
\begin{equation}
\boldsymbol{\Sigma}_i = \mathbf{R}_i \mathbf{S}_i \mathbf{S}_i^T \mathbf{R}_i^T
\label{eq:covariance}
\end{equation}
where $\mathbf{S}_i$ is a diagonal matrix with the scaling factors $\mathbf{s}_i$ on the diagonal, and $\mathbf{R}_i$ is the $3 \times 3$ rotation matrix derived from the quaternion $\mathbf{q}_i$. Note that $\mathbf{S}_i^T = \mathbf{S}_i$ since it's diagonal, so $\boldsymbol{\Sigma}_i = \mathbf{R}_i \mathbf{S}_i^2 \mathbf{R}_i^T$. The anisotropic nature allows the representation to capture effects that have distinct spatial extents and orientations.

 \subsection{Path Loss Prediction Model}

Given a transmitter $\mathbf{tx}$ and receiver $\mathbf{rx}$, the path loss offset $\Delta PL_{\text{model}}$ is calculated by summing the contributions of all Gaussians $G_i$ based on their interaction with the path segment connecting $\mathbf{tx}$ and $\mathbf{rx}$.

In principle, this interaction could involve integrating the Gaussian influence along the path. However, inspired by the projection approach in 3DGS rendering and for computational efficiency, we approximate the influence of a Gaussian $G_i$ on the link $(\mathbf{tx}, \mathbf{rx})$ based on the proximity of the Tx--Rx path segment to the Gaussian's center $\boldsymbol{\mu}_i$.

Let the Tx--Rx path segment be parameterized by arc length
\(\ell \in [0,d]\) as
\[
\mathbf{p}(\ell) = \mathbf{tx} + \ell\,\mathbf{u},
\]
where \(\mathbf{v}=\mathbf{rx}-\mathbf{tx}\), \(d=\|\mathbf{v}\|_2\), and
\(\mathbf{u}=\mathbf{v}/d\) is the unit direction from Tx to Rx. For a Gaussian
with mean \(\boldsymbol{\mu}_i\), define \(\mathbf{w}_i=\boldsymbol{\mu}_i-\mathbf{tx}\).
Projecting \(\mathbf{w}_i\) onto \(\mathbf{u}\) yields the signed distance (in meters)
from \(\mathbf{tx}\) to the closest point on the infinite line:
\begin{equation}
\ell_{\text{proj}, i} = \mathbf{w}_i \cdot \mathbf{u}.
\label{eq:lproj}
\end{equation}
The closest point on the infinite line is then
\begin{equation}
\mathbf{p}_{\text{line}, i} = \mathbf{tx} + \ell_{\text{proj}, i}\,\mathbf{u}.
\label{eq:pline}
\end{equation}

A Gaussian \(G_i\) is considered relevant to the link segment \((\mathbf{tx},\mathbf{rx})\)
if this closest point lies within the segment endpoints, i.e.,
\(0 < \ell_{\text{proj}, i} < d\). Let \(\mathcal{I}(\mathbf{tx}, \mathbf{rx})\) denote
the set of indices of such relevant Gaussians. For \(i\in\mathcal{I}(\mathbf{tx}, \mathbf{rx})\),
the perpendicular displacement from the Gaussian mean to the link is
\begin{equation}
\boldsymbol{\delta}_i = \mathbf{p}_{\text{line}, i} - \boldsymbol{\mu}_i.
\label{eq:delta}
\end{equation}
By construction, \(\boldsymbol{\delta}_i\) is orthogonal to the path direction \(\mathbf{u}\),
and \(\|\boldsymbol{\delta}_i\|_2\) is the perpendicular distance from \(\boldsymbol{\mu}_i\)
to the Tx--Rx segment.

The influence $\alpha_i$ of a relevant Gaussian $G_i$ on the link is modeled using a function that decays with the distance from the path to the Gaussian center, considering the Gaussian's shape and orientation (covariance $\boldsymbol{\Sigma}_i$). This is analogous to evaluating an unnormalized Gaussian probability density, but based on the distance vector $\boldsymbol{\delta}_i$:
\begin{equation}
\alpha_i(\mathbf{tx}, \mathbf{rx}) = \exp\left( -\frac{1}{2} \boldsymbol{\delta}_i^T \boldsymbol{\Sigma}_i^{-1} \boldsymbol{\delta}_i \right) \quad \text{if } 0 < \ell_{\text{proj}, i} < d
\label{eq:influence_mahalanobis}
\end{equation}
and $\alpha_i(\mathbf{tx}, \mathbf{rx}) = 0$ otherwise. Computationally, this Mahalanobis distance is more easily calculated by transforming $\boldsymbol{\delta}_i$ into the Gaussian's local coordinate system using the inverse rotation $\mathbf{R}_i^T$:
\begin{equation}
\boldsymbol{\delta}_{\text{local}, i} = \mathbf{R}_i^T \boldsymbol{\delta}_i = [\delta_{x',i}, \delta_{y',i}, \delta_{z',i}]^T
\end{equation}
Then, the exponent term becomes a sum of squared distances scaled by the learned scaling factors $s_{k,i}$:
\begin{equation}
\alpha_i(\mathbf{tx}, \mathbf{rx}) = \exp\left( -\frac{1}{2} \sum_{k=1}^3 \left(\frac{\delta_{k',i}}{s_{k,i}}\right)^2 \right) \quad \text{if } 0 < \ell_{\text{proj}, i} < d
\label{eq:influence_local}
\end{equation}
This influence factor $\alpha_i$ is close to 1 when the Tx--Rx path passes very near the center $\boldsymbol{\mu}_i$ (relative to the Gaussian's scale $\mathbf{s}_i$) and decays exponentially as the path moves away.

The total path loss offset is then modeled as the sum of the offset contributions $o_i$ from all relevant Gaussians, weighted by their respective influences $\alpha_i$. Let $N$ denote the total number of Gaussians in our model. Then:
\begin{equation}
\Delta PL_{\text{model}}(\mathbf{tx}, \mathbf{rx}) = \sum_{i=1}^N o_i \cdot \alpha_i(\mathbf{tx}, \mathbf{rx}).
\label{eq:pl_deviation_sum}
\end{equation}

The goal is to predict the path loss, PL, between a transmitter at position $Tx \in R^3$ and a receiver at position $Rx \in R^3$ for the operating frequency, $f$ of the training and test measurements. We decompose the total path loss into two components: a baseline path loss model, $PL_{base}$, and a learned offset component, $\Delta PL_{model}$, representing the site-specific effects induced by the environment:

\begin{equation}
PL_{\text{pred}}(\mathbf{tx}, \mathbf{rx}) = PL_{base}(\mathbf{tx}, \mathbf{rx}, f) + \Delta PL_{\text{model}}(\mathbf{tx}, \mathbf{rx}, f)
\label{eq:total_pl}
\end{equation}

The baseline model captures the fundamental distance and frequency-dependent signal attenuation, typically modeled using the Free Space Path Loss (FSPL) formula, but incorporating a learnable path loss exponent (PLE), $\gamma$. Let $d = ||\mathbf{rx} - \mathbf{tx}||_2$ be the Euclidean distance between the Tx and Rx, clamped to a minimum value (e.g., \SI{1}{m}) to avoid singularities. The baseline path loss is:
\begin{equation}
PL_{base}(\mathbf{tx}, \mathbf{rx}, f, \gamma) = 20 \log_{10}(f) + 10\gamma \log_{10}(d) + C
\label{eq:pl_base}
\end{equation}
where $f$ is the frequency in Hz, $d$ is the distance in meters, $C = 20 \log_{10}\left(\frac{4\pi}{c}\right)$ is a constant derived from the FSPL formula with $c$ being the speed of light, and $\gamma$ is the path loss exponent. In our framework, $\gamma$ is treated as a learnable parameter that is initialized to 2.0 and optimized jointly with the Gaussian parameters to best fit the average propagation trend in the measured data for the specific environment. 

\subsection{Initialization Strategy}

The initial placement of Gaussians critically affects convergence. We employ a data-driven initialization:

\begin{equation}
\boldsymbol{\mu}_i^{(0)} = \mathbf{tx}_j + t_i (\mathbf{rx}_j - \mathbf{tx}_j), \quad t_i \sim \mathcal{U}(0.1, 0.9)
\end{equation}

where measurements $j$ are randomly sampled from $\mathcal{D}$. Initial scales are set proportional to the measurement spacing:
\begin{equation}
s_{k,i}^{(0)} = \sigma_0 \cdot \text{median}(\{d_j\}), \quad \sigma_0 \in [0.1, 0.5]
\end{equation}

Quaternions $\mathbf{q}_i^{(0)}$ are initialized to identity (no rotation), and offsets $o_i^{(0)} \sim \mathcal{N}(0, \sigma_o^2)$ with small variance.

\subsection{Optimization and Training}
    An illustration of PropSplat's training cycle is in Fig.~\ref{fig:training}. The parameters of the model, including the Gaussian parameters $\{\boldsymbol{\mu}_i, \mathbf{s}_i, \mathbf{q}_i, o_i\}_{i=1}^N$, and the learnable path loss exponent $\gamma$ are optimized end-to-end using gradient descent. The objective is to minimize the error between the predicted path loss $PL_{\text{pred}}$ (Eq.~\eqref{eq:total_pl}) and the ground truth path loss $PL_{gt}$ from the sparse measurement dataset $\mathcal{D} = \{(\mathbf{tx}_j, \mathbf{rx}_j, PL_{gt, j})\}$.

We use the MSE loss function. To counteract non-uniform sampling density, a distance-based weighting scheme is applied:
\begin{equation}
\mathcal{L} = \frac{1}{|\mathcal{D}|} \sum_{j=1}^{|\mathcal{D}|} w(d_j) \cdot (PL_{\text{pred}}(\mathbf{tx}_j, \mathbf{rx}_j) - PL_{gt, j})^2
\end{equation}
where $d_j = ||\mathbf{rx}_j - \mathbf{tx}_j||_2$ is the distance for the $j$-th sample. The weight $w(d_j)$ is defined as $w(d_j) = (d_j + \epsilon)^p / \overline{(d+\epsilon)^p}$, where $p$ is a hyperparameter and the denominator provides normalization over the batch or dataset. This gives more weight to longer links, which often exhibit larger path loss values and might be underrepresented in dense urban datasets.

\section{Results}

\begin{figure*}[t]
    \centering    \includegraphics[width=0.84\linewidth]{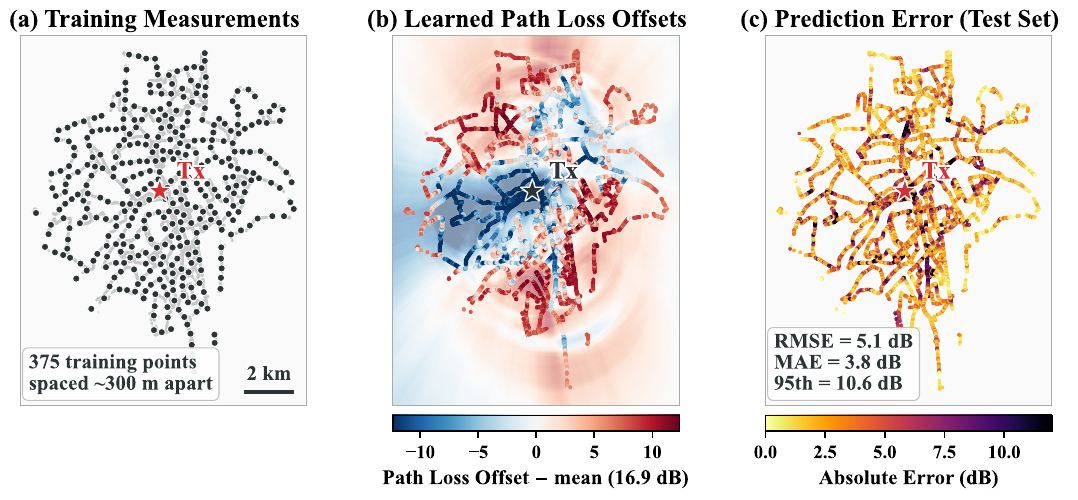}
        \caption{PropSplat applied to Ofcom London 5850~MHz drive-test measurements. Routes span up to 8 km from Tx.
  \textbf{(a)}~Sparse training set with 375 Rx measurements (black dots) spaced ${\approx}$300\,m apart, selected from $>$130{,}000
  drive-test points.
  \textbf{(b)}~Visual comparison of PropSplat's predictions against ground truth. The faded background shows PropSplat's predicted offset
  field. Overlaid points show measured offsets along held-out test routes. Both are relative to the learned baseline (PLE\,$=1.79$) minus
  spatial mean (16.9\,dB). Red\,$=$\,above-average loss, blue\,$=$\,below.
  \textbf{(c)}~Prediction error on held-out test data.}  
 
    \label{fig:london-example}
\end{figure*}

\label{sec:results}

We evaluate PropSplat across two complementary scenarios: large-scale outdoor propagation reconstruction using Ofcom drive-test measurements and indoor BLE localization in a complex multipath environment. Together, these experiments validate that 3D Gaussian primitives can learn RF propagation phenomena across scales and environments without geographic priors.

\subsection{Large-Scale Outdoor Propagation Evaluation}
\label{sec:ofcom}

PropSplat is evaluated on large-scale outdoor measurements spanning diverse topographical environments to test its ability to reconstruct accurate RF fields with ultra-sparse data. This can help reduce drive-test costs and benefit crowdsourced network optimization.

\subsubsection{Dataset Description}
We use Ofcom sub-6 GHz drive-test measurements, collected across seven UK locations: Boston, London, Merthyr Tydfil, Nottingham, Southampton, Stevenage, and Scar Hill~\cite{Ofcom2019Sub6GHz}. These measurements span urban, suburban, rural, flat, and mountainous topographies. Six frequencies were measured at each location: 449, 915, 1802, 2695, 3602, and 5850\,MHz. A continuous wave Tx (mast height \SI{17}{m}; \SI{25}{m} in London) and a vehicle-mounted Rx at \SI{1.5}{m} collected measurements along routes up to \SI{25}{km}, yielding tens of thousands to hundreds of thousands of measurements per city--frequency combination.

\subsubsection{Experimental Setup}
We evaluate PropSplat under sparse training measurements, specifically \SI{300}{m} Rx spacing which is ${\approx}0.8$\% of data. The sparse measurement spacing is visualized for London in Fig.~\ref{fig:london-example}. Dense (80/20 split) training scenarios were also conducted for PropSplat. Each city--frequency pair is run 10 times. Fig.~\ref{fig:propsplat-ofcom-vary-distance} additionally varies the spacing from \SI{100}{m} to \SI{500}{m}. PropSplat is compared against three wireless radiance field methods, NeRF$^2$~\cite{10.1145/3570361.3592527}, GSRF~\cite{yang2025gsrf}, and WRF-GS+~\cite{11258087}, as well as a map-based FCNN~\cite{bocus2025application}, and Kriging. All wireless radiance field methods and Kriging use the identical 300\,m spatial split.

 \subsubsection{Main Results}

Table~\ref{tab:ofcom_summary_metrics} presents the performance comparison. In the sparse training scenario (\SI{300}{m} spacing, $<$1\% of data), PropSplat achieves \SI{5.38}{dB} RMSE averaged across all city--frequency pairs. Among the wireless radiance field baselines, WRF-GS+ is the closest competitor at \SI{5.87}{dB} RMSE, followed by GSRF at \SI{7.46}{dB}. NeRF$^2$ performs poorly at \SI{14.76}{dB} RMSE, nearly $3\times$ worse than PropSplat. Kriging (\SI{9.23}{dB}) falls between these extremes. PropSplat's 95th percentile error of \SI{11.2}{dB} is comparable to WRF-GS+'s \SI{11.9}{dB}, indicating similar worst-case reliability, while GSRF's (\SI{15.1}{dB}) and NeRF$^2$'s (\SI{31.4}{dB}) tails are substantially heavier.

In the dense training scenario (80/20 split), PropSplat achieves \SI{3.08}{dB} RMSE, a 55\% improvement over the map-based FCNN's \SI{6.9}{dB}~\cite{bocus2025application} without using any geographic priors. The FCNN uses clutter and terrain depth information in addition to drive-test measurements.

\begin{figure}[t]
    \centering
    \includegraphics[width=0.94\linewidth]{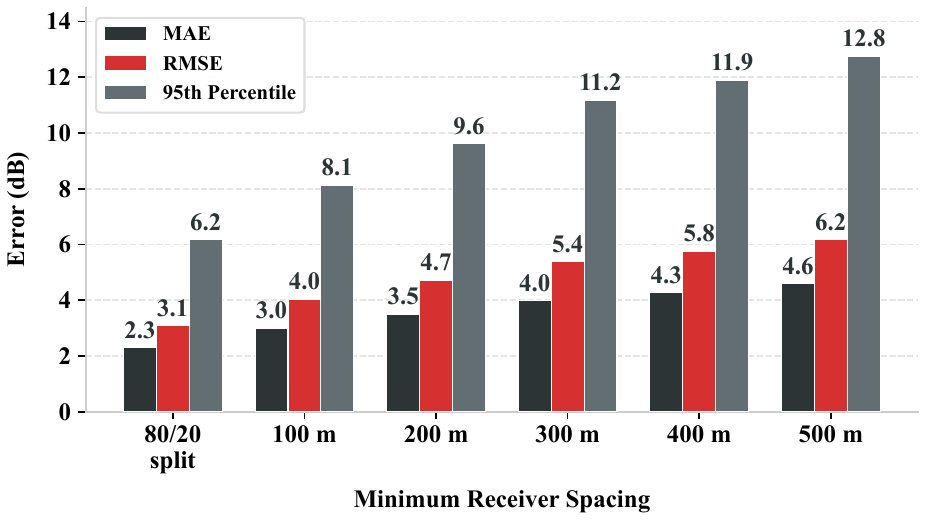}
    \caption{PropSplat performance, varying the measurement distance used for training data between 100 meters and 500 meters. A 80/20 train-test data split is also shown for reference.}
    \label{fig:propsplat-ofcom-vary-distance}
\end{figure}

Fig.~\ref{fig:propsplat-ofcom-vary-distance} shows PropSplat's performance when training measurements vary in spacing from \SI{100}{m} to \SI{500}{m}. Even at \SI{500}{m}, PropSplat remains competitive with map-based methods using dense data. At inference, PropSplat generates 250,000 predictions per second (\SI{0.004}{ms} latency), enabling real-time spectrum management and on-site model updates as drive-tests progress.

\subsubsection{PropSplat Component Ablation}
To quantify the contribution of PropSplat's major components, an ablation study is performed. We toggle both the spatial representation and the path loss backbone. Table~\ref{tab:ablation} reports average RMSE and training time across all city-frequency combinations, displaying how each component affects performance. 

\textbf{Learnable Path Loss Exponent (-LPLE):} Fixing the path loss exponent to 2.0 (free space) forces Gaussians to compensate for environment-specific propagation trends, resulting in a \SI{1.99}{dB} RMSE increase. The learnable PLE adapts to each environment's average propagation characteristics, providing an appropriate baseline for Gaussian offsets to refine. Without this adaptability, Gaussians must model both local variations and global trends, reducing their effectiveness.

\textbf{No Gaussians:} This configuration removes all 3D Gaussians, reducing the model to free-space path loss with a learnable exponent (FSPL+LPLE). The dramatic RMSE increase of \SI{4.44}{dB} (to \SI{9.82}{dB}) demonstrates that Gaussians are essential for capturing spatial propagation variations beyond simple distance-based models. This baseline shows that PropSplat's 3D Gaussian representation provides substantial modeling capability beyond traditional analytical models.

\begin{table}[t]
\centering
\caption{Overall performance on the Ofcom dataset, averaged across all cities and frequencies.}
\label{tab:ofcom_summary_metrics}
\begin{threeparttable}
\resizebox{\columnwidth}{!}{%
\begin{tabular}{@{}lccc@{}}
\toprule
\textbf{Method} & \textbf{MAE (dB)} & \textbf{RMSE (dB)} & \textbf{95th Pct. Err. (dB)} \\ \midrule
\multicolumn{4}{@{}l}{\textbf{Dense Training (80/20 split)}} \\
PropSplat                                & \textbf{2.30} & \textbf{3.08} & \textbf{6.17}  \\
FCNN~\cite{bocus2025application}$^{\dagger}$ & —    & 6.90 & —     \\
\midrule
\multicolumn{4}{@{}l}{\textbf{Sparse Training ($\approx$\,300\,m spacing, $<$1\% data)}} \\
PropSplat                                & \textbf{3.98} & \textbf{5.38} & \textbf{11.17} \\
WRF-GS+~\cite{11258087}                 & 4.34 & 5.87 & 11.89 \\
GSRF~\cite{yang2025gsrf}                 & 5.66 & 7.46 & 15.11 \\
Kriging$^{\beta}$                        & 7.13 & 9.23 & 20.68 \\
NeRF$^2$~\cite{10.1145/3570361.3592527} & 10.14 & 14.76 & 31.37 \\
\bottomrule
\end{tabular}%
}
\begin{tablenotes}\footnotesize
\item[$\dagger$] Uses map data (clutter and terrain depth).
\item[$\beta$] Kriging implemented as Gaussian process regression.
\end{tablenotes}
\end{threeparttable}
\end{table}

\textbf{Anisotropic Gaussians (-ANISO):} Constraining Gaussians to isotropic (spherical) shapes limits their expressiveness. The \SI{1.78}{dB} RMSE degradation highlights how anisotropic shapes capture directional propagation effects.

\subsubsection{Discussion}

The large-scale outdoor results show two performance tiers. NeRF$^2$ (\SI{14.76}{dB} RMSE) performs nearly $3\times$ worse than the Gaussian-based methods, which range from PropSplat (\SI{5.38}{dB}) to GSRF (\SI{7.46}{dB}). PropSplat's ablation (Table~\ref{tab:ablation}) and an analysis of each method's forward-pass structure provide evidence for what drives these differences.

\textbf{Contribution independence.} The methods differ in how individual elements contribute to a prediction. PropSplat computes a weighted sum of independent scalar offsets. Removing any single Gaussian changes the prediction by exactly that Gaussian's contribution (0\% additivity error), and every active Gaussian maintains a consistent sign where it is always adding or always subtracting loss regardless of query. These properties are architectural and they follow from the weighted-sum structure of the forward pass and hold at any scene scale.

GSRF, WRF-GS+, and NeRF$^2$ use rendering equations that couple element contributions. Each uses transmittance products where removing one element changes how much every subsequent element along the ray contributes. We measure 87\% additivity error for GSRF and 121\% for NeRF$^2$, with NeRF$^2$ additionally showing 0\% sign consistency---every sample point's effective contribution depends on and changes sign with context. These coupling properties are also architectural and scale-invariant.

Path loss in dB is conventionally modeled as additive where obstructions along a propagation path contribute independent dB losses. PropSplat's forward pass mirrors this structure and each Gaussian adds an independent dB offset along the Tx--Rx path. The coupled rendering in the competing methods does not match this additive structure, requiring the optimizer to learn the correct prediction through interacting elements rather than independent ones. 

\textbf{Learnable baseline.} PropSplat's learnable PLE baseline contributes \SI{1.99}{dB} (Table~\ref{tab:ablation}). Neither NeRF$^2$, GSRF, nor WRF-GS+ include a learnable baseline in their original architectures. Evaluating GSRF and WRF-GS+ with a fixed PLE\,$=$\,2.0 baseline yields \SI{7.46}{dB} and \SI{5.87}{dB}, essentially unchanged from variants with a learnable PLE (\SI{7.35}{dB} and \SI{5.71}{dB}). The learnable PLE is critical for PropSplat because its independent scalar offsets cannot represent distance-dependent trends. However, the learnable PLE is redundant for methods whose coupled representations absorb distance dependence implicitly.

\textbf{Gaussian spatial capacity.} Removing all Gaussians raises RMSE by \SI{4.44}{dB} (to \SI{9.82}{dB}, comparable to Kriging at \SI{9.23}{dB}). Anisotropic shapes contribute a further \SI{1.78}{dB}. NeRF$^2$'s \SI{14.76}{dB} is worse than even PropSplat's PLE-only ablation (\SI{9.82}{dB}), confirming that the implicit MLP struggles at city scale beyond the lack of a baseline.

\textbf{Finding.}
PropSplat's Gaussians contribute independently and additively to predictions. This is an architectural property that mirrors the additive-dB structure of path loss. The competing methods couple element contributions through transmittance products that do not match this structure. Combined with a learnable PLE baseline (\SI{1.99}{dB} contribution), PropSplat achieves \SI{5.38}{dB} RMSE from $<$1\% of data and requires no geographic priors.

\subsection{Indoor BLE Signal Modeling and Localization}

PropSplat is evaluated on a Bluetooth Low Energy (BLE) dataset, demonstrating its effectiveness in two tasks: RSSI prediction at fixed Rxs and fingerprint-based localization of mobile Txs.

\subsubsection{Dataset Description}
We use the public BLE dataset introduced by NeRF$^2$~\cite{10.1145/3570361.3592527}. This dataset contains RSSI measurements collected in a \SI{15000}{ft^2} elderly nursing home, a complex indoor environment with multiple rooms, walls, and furniture that create challenging multipath propagation and non-line-of-sight conditions. The dataset contains 21 fixed BLE gateways (GWs) recording RSSI values from mobile BLE beacons at 6,000 Tx locations.

\subsubsection{Experimental Setup}

\begin{table}[t]
  \centering
  \caption{Ablation of PropSplat components averaged across all Ofcom experiments.}
  \label{tab:ablation}
  \setlength{\tabcolsep}{6pt}
  \begin{tabular}{lccccc}
    \toprule
    \textbf{Model variant} & \textbf{RMSE [dB]} & $\Delta$\textbf{RMSE} & \textbf{Time [s]} & $\Delta$\textbf{Time} \\
    \midrule
    \textbf{PropSplat}     & 5.38 & — & 189  & — \\
    \hline
    \multicolumn{5}{l}{\textit{Spatial Representation}} \\
    No Gaussians$^{\dagger}$ & 9.82  & +4.44 & 67 & -122 \\
    -ANISO         & 7.16 &  +1.78 &  79 & -110 \\
    \hline
    \multicolumn{5}{l}{\textit{Path Loss Backbone}} \\
    -LPLE           &  7.37&  +1.99 & 199 & +10 \\
    \bottomrule
  \end{tabular}
  \begin{tablenotes}
  \footnotesize
  \item $^\dagger$ Equivalent to FSPL with learnable path loss exponent only.
  \end{tablenotes}
\end{table}

We compare PropSplat against three wireless radiance field baselines: NeRF$^2$~\cite{10.1145/3570361.3592527}, GSRF~\cite{yang2025gsrf}, and WRF-GS+~\cite{11258087}. All methods are evaluated under identical data splits on two training scenarios. The dense scenario uses a random 70\%/30\% train-test split (4,200 training, 1,800 test positions). The sparse scenario uses 758 training positions (12.6\% of the dataset) with the remaining 5,242 positions for testing. For localization, each trained model generates synthetic RSSI fingerprints over a dense grid, and K-Nearest Neighbors ($K=5$) matching estimates Tx positions.

\subsubsection{Results}
Table~\ref{tab:comparison_results} presents RSSI prediction and localization results. For dense RSSI, PropSplat and NeRF$^2$ achieve near-identical accuracy (\SI{4.74}{dB} vs.\ \SI{4.68}{dB} RMSE), both outperforming GSRF (\SI{5.15}{dB}) and WRF-GS+ (\SI{5.30}{dB}). For sparse RSSI, PropSplat leads all methods with \SI{6.51}{dB} RMSE, a 9\% improvement over NeRF$^2$ (\SI{7.17}{dB}), while the remaining baselines cluster in the \SIrange{7.2}{7.6}{dB} range.

Localization reveals striking differences. PropSplat achieves a median localization error of \SI{0.10}{m} (dense) and \SI{0.25}{m} (sparse), maintaining sub-meter accuracy in both regimes. GSRF and WRF-GS+ achieve reasonable dense medians (\SI{0.27}{m} and \SI{0.21}{m}) but degrade $2$--$3\times$ more than PropSplat under sparse training. NeRF$^2$ achieves a competitive dense median of \SI{0.29}{m} but its mean of \SI{1.84}{m} reveals a heavy tail, and under sparse training it degrades to a mean of \SI{7.33}{m}.

\subsubsection{Discussion}

\begin{figure}[t]

    \centering
    \includegraphics[width=1\linewidth]{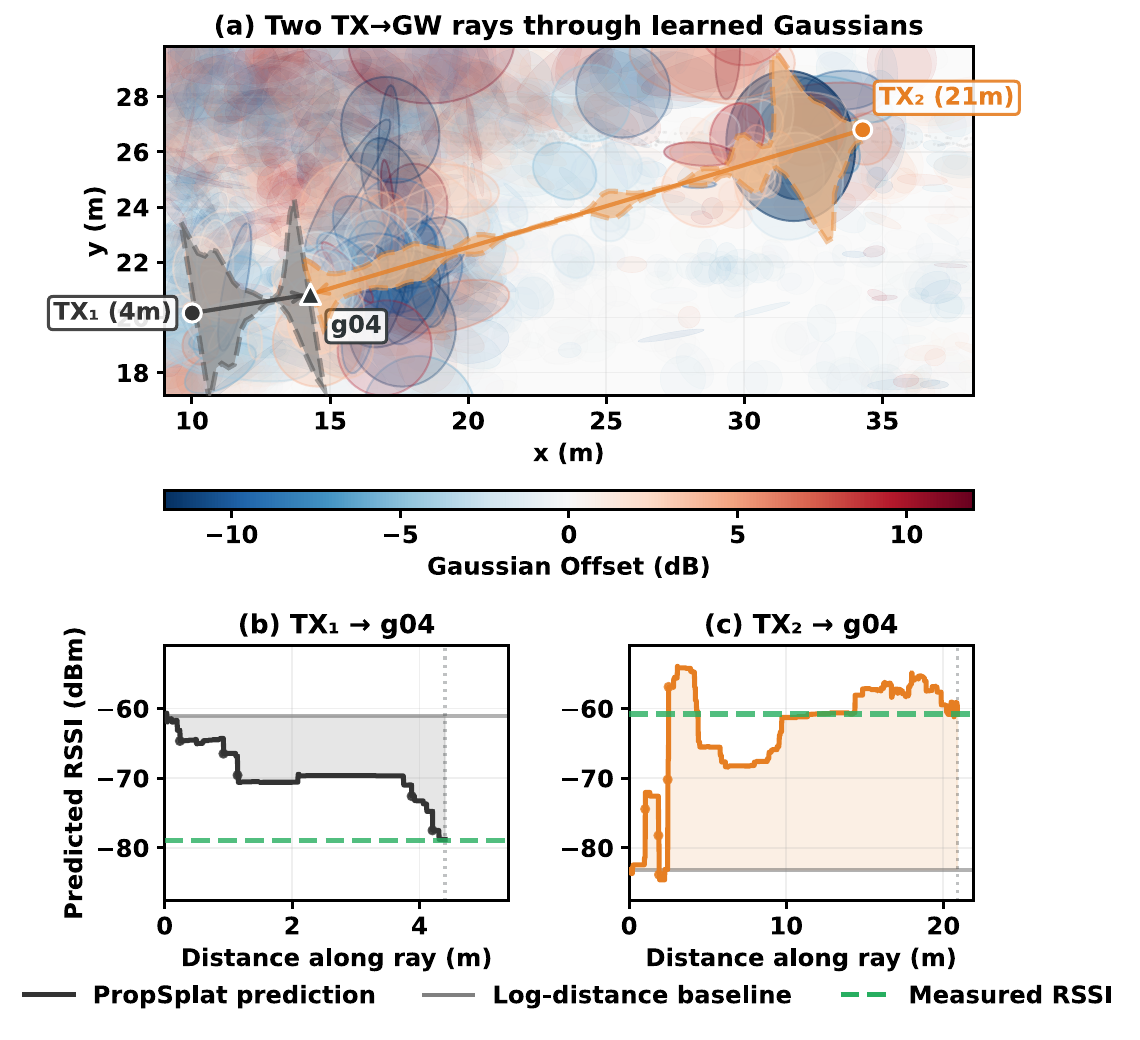}
    \caption{PropSplat's path loss prediction is spatially decomposable where each Gaussian's contribution is localized along the propagation path and individually traceable. Panel \textbf{(a)} shows rays passing through learned Gaussians for two Tx--GW paths in the BLE indoor environment. Gaussians are colored by path loss offset (red: additional attenuation, blue: reduced loss). Both \textbf{(b)} and \textbf{(c)} show cumulative RSSI along each ray as Gaussians are encountered. TX\textsubscript{1} accumulates 17.7\,dB of additional loss while TX\textsubscript{2} receives $-$22.3\,dB of correction. These opposite adjustments capture location-dependent propagation effects. Both match measured RSSI within 0.1\,dB.}
    \label{fig:ble-localization}
\end{figure}

The BLE results show that RSSI prediction accuracy and localization accuracy are not coupled and fundamentally different metrics. All four methods achieve RSSI RMSE within \SI{1}{dB} of each other (\SIrange{4.68}{5.30}{dB} dense), yet localization diverges by an order of magnitude. PropSplat achieves \SI{0.19}{m} mean error while NeRF$^2$ achieves \SI{1.84}{m}.

This decoupling occurs because fingerprint localization matches a 21-dimensional RSSI vector (one per GW) against a dense grid of predictions. Localization succeeds when nearby positions produce distinct fingerprints. It requires spatial discriminativeness, not just low average error. A method can achieve low per-GW RSSI error while producing spatially smooth fields that map many positions to nearly identical fingerprints.

Two observations from the data support this. First, PropSplat localizes at \SI{0.25}{m} median under sparse training where measurements are spaced ${\sim}$\SI{1.2}{m} apart---$5\times$ finer than the training density. This demonstrates that the learned field generalizes spatial structure between training points rather than memorizing them. Second, NeRF$^2$'s localization degrades from \SI{1.84}{m} (dense) to \SI{7.33}{m} (sparse), while PropSplat degrades from \SI{0.19}{m} to only \SI{0.44}{m}. PropSplat's localization is robust to data reduction while NeRF$^2$'s collapses.

The contribution independence property identified in the outdoor analysis (Sec.~\ref{sec:ofcom}) provides a structural explanation for this localization gap. Each of PropSplat's Gaussians imposes a fixed, sign-consistent spatial correction. A small change in position changes which Gaussians contribute and by how much, producing distinct predictions at nearby locations. In contrast, NeRF$^2$'s coupled rendering means that each point's contribution depends on all other points along the ray. Small spatial shifts can produce unpredictable changes in relative contributions, yielding smooth aggregate predictions that lack per-position distinctiveness. Figure~\ref{fig:ble-localization} visualizes PropSplat's spatial structure where each Gaussian's contribution is localized along the propagation path and independently traceable.

\textbf{Finding.}
RSSI accuracy and localization accuracy are decoupled. All methods achieve ${\sim}$\SI{5}{dB} RSSI RMSE, but PropSplat localizes $10\times$ better than NeRF$^2$ and $2$--$3\times$ better than GSRF and WRF-GS+. PropSplat's independent, sign-consistent Gaussian contributions produce spatially discriminative predictions, enabling localization at $5\times$ finer resolution than the training spacing, even at 12.6\% training density.

\section{Limitations and Future Work}
\label{sec:limits-future-work}

PropSplat advances map-free propagation modeling and this section discusses its limitations. The current representation initializes a large number of Gaussians to capture a site's path loss offsets. All experiments are initialized with 9,000 optimizable Gaussians. Each Gaussian contains 11 learnable parameters, creating a high-dimensional optimization problem. While empirically stable, PropSplat could still benefit from a more principled initialization strategy. However, our current approach of initializing Gaussians along Tx--Rx paths provides sufficient constraint for stable optimization. This approach captures the area along and near the location where RF interactions occur, constraining the search space to Gaussians in positions that can immediately contribute to path loss offsets. Without this initialization, the optimizer would have to discover where environmental interactions occur and their corresponding effects simultaneously, across a large-scale space of possible 3D positions. 

PropSplat currently does not account for temporal variations, frequency-selective fading, or multi-antenna system characteristics. Future research directions could address these avenues. Extending PropSplat to model time-varying channels could enable tracking of environmental changes.

\begin{table}[t]
\footnotesize
\centering
\caption{Comparison of BLE Experiment Results.}
\label{tab:comparison_results}
\begin{tabular}{@{}lccc|cc@{}}
\toprule
{Method} & \multicolumn{3}{c|}{RSSI Error (dB)} & \multicolumn{2}{c}{Localization Error (m)} \\
                        & RMSE & MAE & Med. & Mean & Med. \\
\midrule
\multicolumn{6}{@{}l}{\textbf{Sparse Training (758 positions, 12.6\%)}} \\
GSRF~\cite{yang2025gsrf}                & 7.38 & 5.33 & 3.85 & 1.00 & 0.56 \\
WRF-GS+~\cite{11258087}                & 7.56 & 5.24 & 3.58 & 1.65 & 0.67 \\
NeRF$^2$~\cite{10.1145/3570361.3592527} & 7.17 & 5.30 & 3.99 & 7.33 & 3.57 \\
PropSplat                               & \textbf{6.51} & \textbf{4.67} & \textbf{3.44} & \textbf{0.44} & \textbf{0.25} \\
\midrule
\multicolumn{6}{@{}l}{\textbf{Dense Training (70/30 train-test split)}} \\
GSRF                                    & 5.15 & 3.74 & 2.97 & 0.50 & 0.27 \\
WRF-GS+                                & 5.30 & 3.72 & 2.88 & 0.45 & 0.21 \\
NeRF$^2$                                & \textbf{4.68} & \textbf{3.50} & 2.91 & 1.84 & 0.29 \\
PropSplat                               & 4.74 & 3.55 & \textbf{2.88} & \textbf{0.19} & \textbf{0.10} \\
\bottomrule
\end{tabular}
\end{table}

\section{Conclusion}
\label{sec:conclusion}

This work presents PropSplat, a site-specific propagation model that learns from sparse wireless measurements without map data. On the Ofcom outdoor dataset, PropSplat achieves \SI{5.38}{dB} RMSE using $<$1\% of available drive-test data, outperforming all wireless radiance field baselines and a map-based neural network. In indoor BLE experiments, it achieves \SI{0.19}{m} mean localization error, an order of magnitude better than NeRF$^2$, while matching RSSI prediction accuracy.

Both evaluations reveal a consistent architectural pattern. NeRF$^2$'s implicit MLP achieves competitive RSSI accuracy indoors but degrades to \SI{14.76}{dB} RMSE at outdoor scale. Gaussian-based methods (GSRF at \SI{7.46}{dB}, WRF-GS+ at \SI{5.87}{dB}) scale better through explicit spatial representations, but use coupled rendering equations with transmittance products that make each Gaussian's contribution dependent on all others along the ray. PropSplat's forward pass is additive where each Gaussian contributes an independent, sign-consistent scalar offset, mirroring the additive-dB structure of path loss. This independence, combined with a learnable PLE baseline that analytically removes the dominant distance trend, produces predictions that are both accurate in aggregate and spatially discriminative. Overall, PropSplat's architecture enables sub-meter localization indoors and reliable coverage reconstruction outdoors.

 \bibliographystyle{IEEEtran}
\bibliography{ref}

\end{document}